\documentstyle[eqsecnum,preprint,aps]{revtex}
\begin{document}
\draft
\begin{title}
{\Large \bf Quantum Monte
Carlo simulations of a particle in a random potential.}
\end{title}
\author{Hsuan-Yi Chen and Yadin Y. Goldschmidt\footnote{
On sabbatical leave at the Weizmann Institute of Science, Rehovot Israel}}
\address{Department of Physics and Astronomy\\
University of Pittsburgh\\
Pittsburgh, PA 15260}
\date{9 August 1996}
\maketitle

\begin{abstract}
In this paper we carry out Quantum Monte Carlo simulations of a quantum
particle in a one-dimensional random potential (plus a fixed harmonic
potential) at a finite 
temperature. This is the simplest model of an interface in a disordered
medium and may also pertain to an electron in a dirty metal. We compare with
previous analytical results, and also derive an expression for the sample to
sample fluctuations of the mean square displacement from the origin which is
a measure of the glassiness of the system. This quantity as well as the mean
square displacement of the particle are measured in the simulation. The
similarity to the quantum spin glass in a transverse field is noted. The
effect of quantum fluctuations on the glassy behavior is discussed.
\end{abstract}


\newpage

\section{Introduction}

The problem of a quantum spin glass in a transverse field was recently in
the center of theoretical and experimental interest \cite
{yglai,wu,sachdev,mft,young}. In particular the question of the interplay
between glassy behavior and quantum fluctuations and the properties of the
quantum transition at zero temperature were the subject of several
investigations, both for the case of the infinite-ranged spin glass and more
realistic three dimensional models.

A simpler model which catches many of the essential features of the
(classical) spin glass problem is that of a directed manifold in a random
medium \cite{villain,henley,kardar,mp1,yg1}. An even higher simplification
occurs for the case of a zero dimensional manifold, which is equivalent to a
particle in a random potential (which is localized by an additional fixed
harmonic potential) \cite{villain2,brezin,mp,gb1,engel,yg2}. This model has
been found to require a (infinite-step) Parisi type solution when the random
potential has long range correlations \cite{mp}. For a random potential with
short ranged correlations a one-step replica-symmetry-breaking (RSB)
solution has been found\cite{mp1,engel,yg2}. A single particle in one
dimension does not have a sharp transition into a glassy phase. But in
infinite dimensions it does. It turns out that the analytical solution which
utilizes the variational approximation, still possess a sharp transition at
finite dimensions (including one dimension). This occurs since the replica
symmetric (RS) is not able to capture the glassy features of the systems
once they become strong enough, and gives rise to unphysical results, like a
non-monotonic mean square displacement of the particle from the origin as a
function of the temperature. Below a certain temperature the RSB solution
yields a much better physical result, and in particular the correct
non-analytic behavior at $T=0$ as a function of the strength of the random
potential in agreement with an Imry-Ma type argument \cite{engel}. One
should notice though, that the transition which has been found for a
particle in a random potential is of the Almeida-Thouless type \cite{at,b-y}
in the sense that it is associated with RS- RSB transition but not with an
order-disorder (spin-glass like) transition.

The success of the variational treatment of a classical particle in a random
potential, led one of us recently \cite{yyg1,yyg2} to investigate a quantum
analog, i.e. to turn on $\hbar$ and consider the effect of quantum
fluctuations, e.g. tunneling on the glassy behavior of the particle. This
was suggested by the recent theoretical treatment of the quantum spin glass
in a transverse field \cite{yglai,mft}.

Analytically we have found a glassy phase characterized by RSB, which is
destroyed by the quantum fluctuations for strong enough $\hbar $, or
alternatively for small enough $m$, which is the particle's mass. The
variable $\hbar ^2/m$ plays the role of the transverse field in the quantum
spin glass problem. A schematic phase diagram is depicted in Fig. 1. The
full details of the analytical investigation are given in \cite{yyg2} , to
be referred in the sequel as I.

Recently \cite{yyg3} one of us used the model of a particle in two spatial 
dimensions under the influence of an harmonic and a quenched random potential 
to describe the melting transition of the flux lattice in high-temperature
superconductors with columnar disorder. The so called cage model has been 
introduced originally by Nelson and Vinokur \cite{NV}. In this model a 
single flux line is represented by the world line of a quantum particle.
The influence of neighboring flux lines is taken effectively as the cage 
harmonic potential. The magnitude of $\hbar$ is determined by the size
of the system along the z-axis and the value of the temperature. This
shows the usefulness of the toy-model to other physical systems of interest.

Our aim in this paper is two-fold. First we carried out a quantum Monte
Carlo simulation of the system in one dimension in order to compare
with the analytical 
results obtained in I for the mean square displacement. Second, since both
the random potential and the quantum fluctuations increase the mean square
displacement, this quantity by itself is not enough to give a clear picture
concerning the strength of the glassy behavior of the system. Hence we have
measured in the simulation, and also calculated analytically the sample to
sample fluctuation of $\langle {\bf x}^2\rangle $, which shows that the
glassy behavior of the system diminishes as the quantum fluctuation increase
until the eventual demise of the glassy phase for strong enough
$\hbar^2/m$. This decline in the sample to sample fluctuations is a
gradual effect, which analytically culminates in the 
transition from an RSB solution to a RS solution. As mentioned
previously, in a simulation which is carried out in one dimension, we
do not expect to observe any sharp transition. The trapping of the 
particle in deep local minima of the random potential gives rise to a sticky
behavior- i.e. a freezing of the mean square displacement from the origin.
This effect is countered by tunneling among the different minima which
enables the particle to escape from a local minima and thus diminishes the
glassy behavior.

In the next section we define the model. In Section 3 we review some of the
results obtained in I and obtain a new analytic result for the sample to
sample fluctuations of the mean square displacement of the particle from the
origin. In Sect. 4, we describe the details of the quantum Monte Carlo (QMC)
simulation. Section 5 is devoted to a discussion of the results and a
summary. A comparison is made between the simulation and the theoretical
results obtained in I and in Section 3. In the Appendix we give some further
details of the calculation presented in Section 3.

\section{The Model}

The Partition function for a particle at finite temperature $T=1/k_B\beta $,
subject to a harmonic potential and a random potential $V$, is given by the
functional integral \cite{feynman}: 
\begin{eqnarray}
Z(U)\ =\ \int_{{\bf x}(0)={\bf x}(U)}^{}[d{\bf x}]\ \exp \left\{ -{\frac 1%
\hbar }\int_0^U\left[ {\frac{m\dot {{\bf x}}(u)^2}2}\ +\ {\frac{\mu {\bf x}%
(u)^2}2}\ +\ V({\bf x}(u))\right] du\right\} ,  \label{PF}
\end{eqnarray}
where ${\bf x}$ is a $N$-dimensional vector ($N$ is the number of spatial
dimensions), and $U=\beta \hbar $. The variable $u$ has dimensions of time
and is often referred to as the Trotter dimension. We observe that the
trajectory ${\bf x}(u)$ forms a closed path. In this paper we are concerned
with a random quenched potential $V(x)$, which is Gaussian distributed. This
means that the probability for a given realization of the potential is given
by: 
\begin{eqnarray}
P(V(x))=C\exp (-\int dxdx^{\prime }V(x)\Delta (x-x^{\prime })V(x^{\prime })),
\label{vgauss}
\end{eqnarray}
with some known function $\Delta (x-x^{\prime })$. It is thus sufficient to
know only the first two moments of the distribution vis. 
\begin{eqnarray}
\left\langle V({\bf x})\right\rangle _R=0,\ \left\langle V({\bf x})V({\bf %
x^{\prime }})\right\rangle _R=-Nf\left( {\frac{({\bf x}-{\bf x^{\prime }})^2}%
N}\right) .  \label{potav}
\end{eqnarray}
where the functions $f$ and $\Delta $ are related to each other. The
function $f$ describes the correlations of the random potential. In this
paper we consider two cases. One, which we call the case of long ranged
correlations of the potential, for which $f$ is taken to decay as a power at
large distances: 
\begin{eqnarray}
f(y)\ =\ {\frac g{2(1-\gamma )}}\ \ (a_0\ +\ y)^{\ 1-\gamma },  \label{f1}
\end{eqnarray}
with $\gamma =1/2$ . This case corresponds in one dimension to an interface
in the random field Ising model \cite{villain} \cite{mp1,gb1,engel}. The
parameter $a_0$ plays the role of a short-distance regulator for $f$. 
Another type of random potential we consider has Gaussian correlations and
we refer to it as the case of short-ranged correlations. For this case, the
function $f$ is taken to decay as a exponential function at large distances: 
\begin{eqnarray}
f(y)\ =\ {\frac g2}\ \ \exp (-{\frac 12}y)  \label{f2}
\end{eqnarray}
In the classical case it has been shown that (for $N=1$) even when
correlations fall exponentially fast, the physics is equivalent within the
variational approximation to the case of random potential with power law
correlations $\ $(eq. (\ref{f1})) and $\gamma =3/2$ at large distances \cite
{mp1,engel}. This fact also holds in the quantum case, as has been
demonstrated in I. Thus we will compare the results obtained in the QMC for
the distribution (\ref{f2}) with results obtained in I for the case of $\gamma
=3/2$.

\section{Theoretical Considerations}

In paper I we investigated the problem using the replica method and the
variational approximation. Here we review briefly some of the formalism and
discuss the new derivation of the sample to sample fluctuation of the mean
square displacement that is a measure of the glassiness of the system and is
needed to compare with the results of the QMC simulations. Some details are
deferred to the Appendix. Readers who are interested only in the details of
the QMC simulation can skip this section.

In the replica method variational approach the system is represented by an $%
n $-body variational Hamiltonian: 
\begin{eqnarray}
h_n\ &=&\ {\frac 12}\int_0^Udu\ \sum_a\left[ m{\bf \dot x}_a^2(u)+\ \mu {\bf %
x}_a^2(u)\right] \hspace{1.75in}  \nonumber \\
&&\ -\ {\frac 1{2\hbar }}\int_0^Udu\int_0^Udu^{\prime
}\sum_{ab}s_{ab}(u-u^{\prime }){\bf x}_a(u)\cdot {\bf x}_b(u^{\prime }).
\label{hn}
\end{eqnarray}
The matrix $s_{ab}(u-u^{\prime })$ is determined by extremizing the
variational free-energy which is given by: 
\begin{eqnarray}
n\beta \left\langle F\right\rangle _R/N\ =\ \langle {\cal H}_n-h_n\rangle
_{h_n}/\hbar -\ln \int [dx]e^{-h_n/\hbar }.  \label{fvar}
\end{eqnarray}
Here ${\cal H}_n$ is the exact $n$-body Hamiltonian. The limit $n\rightarrow
0$ has to be taken at the end.

The propagator associated with $h_{n\text{ }}$is given in frequency space :

\begin{eqnarray}
G_{ab}(\omega )\equiv ([(m\omega ^2+\mu ){\bf 1}-\tilde s(\omega
)]^{-1})_{ab}.  \label{G}
\end{eqnarray}
$\omega $ is the frequency conjugate to the Trotter time variable $u$, and
takes the values:

\begin{eqnarray}
\omega _l={\frac{2\pi }U}l,\ \ l=0,\pm 1,\pm 2,\cdots \ \ ,  \label{omega}
\end{eqnarray}
and the matrix $\widetilde{s}_{ab}(\omega )$ is related to $s_{ab}(u)$ by:

\begin{equation}
s_{ab}(\zeta )={\frac 1\beta }\sum_{l=-\infty }^\infty \exp (-i\omega _l\
\zeta )\ \tilde s_{ab}(\omega _l).  \label{FT}
\end{equation}

We have found a self-consistent solution to the variational equations where
only the diagonal elements of the matrix $s_{ab}$ are 'time' dependent, and
the off-diagonal elements are independent of the Trotter time. Thus

\begin{eqnarray}
\tilde s_{aa}(\omega ) &=&\tilde s_d(\omega ), \\
\tilde s_{ab}(\omega ,z) &\leftrightarrow &\tilde s(z)\ \delta _{\omega ,0}\
,\ \qquad a\neq b,  \label{matrix}
\end{eqnarray}
where the Parisi parameter $0<z<1$ labels the ``distance'' between replicas
indices $ab$. A similar behavior follows for the propagator matrix $%
G_{ab}(\omega )$ with a similar notation $G_d(\omega )$ and $G(z)$ (These
are the same as the quantities $\widetilde{r}_d(\omega )$ and $\widetilde{r}%
(z)$ used in I) .

The mean square displacement from the origin is given by:

\begin{equation}
\langle \langle {\bf x}^2\rangle \rangle _R/N={\frac 1\beta }\sum_{k=-\infty
}^\infty G_d(\omega _k).  \label{x2}
\end{equation}
This quantity was evaluated in I.

The sample to sample fluctuation of $\langle {\bf x}^2\rangle $ is a measure
of the glassiness of the system. In the replica approach together with
the variational approximation, this quantity is represented by 
\begin{eqnarray}
\left\langle \left\langle {\bf x}^2\right\rangle ^2\right\rangle _R-{%
\left\langle \left\langle {\bf x}^2\right\rangle \right\rangle _R}%
^2=\left\langle {\bf x}_a^2{\bf x}_b^2\right\rangle _{h_n}-\left\langle {\bf %
x}_a^2\right\rangle _{h_n}^2  \label{x22a}
\end{eqnarray}
where $a$ and $b$ are indices for replicas. Follow the notation we used in
I, the sample to sample fluctuation of the mean square displacement becomes: 
\begin{eqnarray}
\frac 1{N^2}\left( \left\langle \left\langle {\bf x}^2\right\rangle
^2\right\rangle _R-{\left\langle \left\langle {\bf x}^2\right\rangle
\right\rangle _R}^2\right) =\frac 2{N\beta ^2}\int_0^1G^2(z)dz,  \label{x22b}
\end{eqnarray}
For a particle in $N$ dimensions it is self averaging in the large $N$
limit but 
not for $N=1$. The degree of non-self-averaging is a measure of the glassy
behavior of the system.

In the Appendix we give details of the numerical evaluation of the sample to
sample fluctuation of $\langle \langle {\bf x}^2\rangle \rangle $ both in
the replica symmetric and in the RSB phases. The calculation is done both
for the case of continuous RSB which occurs for a random potential with long
ranged correlations, and for the case of short-ranged correlated
potential where there is a one-step RSB.
The results are depicted in Fig. 2 and Fig. 3 respectively. We observe that
the glassiness of the system increases with decreasing temperature, but
decreases with increasing $\hbar ^2/m\equiv 1/\kappa $. Recall that
the transition temperature $T_c(\kappa)$ between the RS and RSB phases
decrease with decreasing $\kappa$ as was obtained in I.

Another quantity that could also be used as a measure for the glassiness of
the system, but we did not measure in the QMC simulation is the sample to
sample fluctuation of the susceptibility,

\begin{equation}
\chi =\frac 1N\left( \left\langle {\bf x}^2\right\rangle -\left\langle {\bf x%
}\right\rangle ^2\right) ,  \label{sus}
\end{equation}
which is given by:

\begin{equation}
\left\langle \chi ^2\right\rangle _R-\left\langle \chi \right\rangle _R^2=%
\frac 1{3\beta ^2}\left( 1+\frac 2N\right) \left[ \int_0^1dzG^2(z)-\left(
\int_0^1dzG(z)\right) ^2\right] .  \label{stssus}
\end{equation}

\section{Numerical Simulation}

We applied the path integral Monte Carlo method (PIMC) in one space
dimension to calculate the relevant physical quantities we are interested
in. The partition function of this system at a given temperature is given by 
\begin{eqnarray}
Z(T)=\ \int_{{\bf x}(0)={\bf x}}^{{\bf x}(U)={\bf x^{\prime }}}[d{\bf x}]\
\exp \left\{ -{\frac 1\hbar }\int_0^U\left[ {\frac{m\dot {{\bf x}}(u)^2}2}\
+\ {\frac{\mu {\bf x}(u)^2}2}\ +\ V({\bf x}(u))\right] du\right\} ,
\label{densitym}
\end{eqnarray}

We discretize $x(\tau )$ into $M+1$ points with $x(1)$ equal to $x(M+1)$.
The partition function now becomes an $M$-dimensional ordinary integral,
i.e., 
\begin{eqnarray}
Z(T)=\int \prod_{i=1}^M\frac{dx_i}A\exp \left[ -\sum_{j=1}^M\varepsilon
\times \left[ \frac m2(\frac{(x_{j+1}-x_j)}\varepsilon )^2+\frac \mu 2%
x_j^2+V(x_j)\right] \right] ,
\end{eqnarray}
where $\varepsilon =U/M$, $A=\sqrt{2\pi \hbar \varepsilon /m}$. In this way
we are able to apply the Metropolis method to perform the integration for
the partition function.

In this form the partition function of a quantum particle is similar to the
classical Boltzmann distribution of a polymer ring with $M$ beads under the
influence of an applied harmonic potential and a random potential. The beads
have a harmonic spring interaction between neighboring beads and in addition
each bead feels a combination of a harmonic (w.r.t the origin ) and a random
local potentials. This gives us an intuitive picture of the simulation. In
the simulation one attempts to move each bead in its own turn
and one checks if the ``energy'' ( minus the argument of the exponential)
decreases or increases. Then one actually moves the bead in accordance
to a detailed balance algorithm, e.g. the Metropolis algorithm.

The problem in doing PIMC comes from the fact that for small $\varepsilon $
(or large $M)$ the beads are not easy to move due to a very large spring
constant, thus the acceptance rate is low for a reasonable size move and
convergence is not easy to achieve in a reasonable time. For this reason
many efforts have been made to circumvent the problem \cite{Imada}, \cite
{Berne},\cite{Runge}, \cite{Sprik} . We use the normal mode PIMC to perform
the calculation \cite{Imada}. In our program the motion of the `beads'
includes 2 parts,i.e.,

\begin{itemize}
\item  1. Microscopic movement: we attempt to change the value of each $x_i$
individually. We put $x_i^{\prime }=x_i+dx$ and decide whether $x_i$ should
change or not.

\item  2. Global movement: we consider all $x_i$'s together as a set, and
consider $\ $%
\begin{equation}
x_i^{\prime }=x_i+a_0+\sum_{q=1}^{q_c}a_qsin\left( 2\pi q\frac{i-1}M\right)
,\qquad \forall i=1\cdots M.
\end{equation}
\end{itemize}

In our simulation the total number of points in 'time' axis are
chosen such that more points are used for small $\kappa$ regime. The numbers 
range from 7 to 16.  Although this is not a large number, results from
simulations we carried for a simple harmonic oscillator give pretty small
errors.  This gives us confidence that in the presence of random
potentials the small number of points in the Trotter time dimension will 
also give us a small error in comparison with the statistical error which comes
from the small number of samples of random potentials.
Since this is not a very large number of points we found it
sufficient to include only the zeroth and first normal modes , i.e. we
choose $q_c=1$, and then decide whether they move or not. The magnitude of $%
dx$ and the $a_i$'s are chosen so that the acceptance ratio is approximately
0.5. The parameters $a_0$, $a_1$, and the size of microscopic movement $dx$
are listed in Table 1 at the end of this section.

We have generated two kinds of random potentials: A random potential with
long range correlation, characterized by $\gamma =1/2$ and a random
potential with short range correlation which decay as a Gaussian. Within the
framework of the variational approximation such a potential is equivalent to
a random potential with power law correlations that are characterized by an
index $\gamma =3/2$.

For the case of long range correlation, we have generated $K=6000$
uncorrelated random numbers $h_1\cdots h_K$ with a Gaussian distribution. We
then constructed the variables $V_0\cdots V_K$ by:

\begin{equation}
V_i=const.\times \left( \sum_{j=0}^ih_j-\sum_{j=i}^Kh_j\right)
\end{equation}
and placed them on one dimensional lattice with a lattice constant chosen to
be $0.01$.  The random potential generated in this way has correlation
\begin{eqnarray}
\left\langle V_i V_j \right\rangle _R = - const. \times |i-j|  + C
\end{eqnarray}
where $C$ is a constant independent of $i$, $j$.  The free energy
depends trivially on $C$ but  $\left\langle \left\langle 
{\bf x}^2 \right\rangle \right\rangle _R$ is independent of $C$ as has 
been mentioned in equation (2) of ref\cite{mp}.
In this way a random potential with long range correlation is
generated. These 6000 numbers have long ranged correlations with $\gamma
=1/2.$

We now discuss the procedure to generate a random potential with short
range correlations.  From the fact that random potentials with correlation
\begin{eqnarray}
\left\langle V(x_1)V(x_2) \right\rangle _R \propto \exp(-\frac{a}{2}
\times (x_1-x_2)^{2}) 
\end{eqnarray}
in configuration space have correlations in momentum space with the following
form
\begin{eqnarray}
\left\langle V_{k_1}V_{k_2} \right\rangle _R \propto \delta (k_1+k_2) \times
\exp(-{k_1}^{2}/2a),
\end{eqnarray}
we generated random numbers with a proper distribution in momentum
space and fast Fourier transformed them by a standard Fortran subroutine  
\cite{nr} to get random numbers with a Gaussian {\it correlation}. We put $%
4000$ of them on lattice sites with a lattice constant of $0.005$. These
constitute a random potential with short (Gaussian) correlations.  
In both long and short-ranged disorder, we discretize the $x$ direction
such that the lattice is about 2 orders of magnitude smaller than
$\left\langle \left\langle {\bf x}^2 \right\rangle \right\rangle _R$.

Another difficulty is that in the low temperature regime the relaxation time
of the system is very long because the phase space of the system is
separated by high free energy barriers and it is difficult to get reliable
results within a reasonable computer time. The standard approach of doing a
simulation in these systems is the simulated annealing approach, and in
addition to using this method we also used a modified version of the global
movement algorithm to speed up the dynamics.

We started our simulation from the high temperature regime ($T=3T_c(\infty)$
for long range correlation case and $T=2T_c(\infty)$ for short range 
correlaion case). 
 We then lowered the temperature in steps of $
1T_c(\infty)$ when $T>T_c(\infty)$. For $T<T_c(\infty)$ we lowered the
temperature by $\delta T=0.1T_c(\infty)$ at a time and performed
thermal averages for every $0.2T_c(\infty)$. The 
lowest temperature of our simulation was set to be $0.2T_c(\infty)$.
In addition, we 
attempted 2 different size (zeroth mode) global movements in each sweep, of
magnitudes $a_0$ and $a_0^{\prime }$ respectively. Generally we chose one of
these parameters, say $a_0$, to be much bigger than the other (but such the
acceptance rate will not fall below $0.01$ ) in order that the particle will
get a chance to occasionally escape from deep wells which correspond to
metastable minima.  We have performed the simulation for different
given random potentials with thermalizaion sweeps ranging from 30,000
to 200,000 and averaged over 200,000 to 2,000,000 MC steps and only
very small difference have been found.  When taking data we have
chosen the number of thermalizaion sweeps to be 50,000 for all cases and 
averaged over 200,000 to 400,000 steps with more steps for lower temperature.

Table 1 gives an example for the parameters we have chosen for the 
simulation in the case of long ranged correlated potential.

\section{Results and Discussion}

The mean square displacement of the particle as well as the sample to sample
fluctuations of the mean square displacement are calculated in our
simulation, and all data points are averaged over 2000 samples of the random
potential.

For the classical case, previous numerical results for the long range
case, were reported by \cite{mp} for the long ranged case and also by \cite
{engel} for both the short and long ranged cases. They did not perform
Monte Carlo simulation, but used the fact that in the classical case instead
of a path integral one has to evaluate a simple integral for each
realization of the random potential. For large $\kappa (=100)$, where the
results should reduce to the classical case, we have checked our results for
the mean square displacement against their's, and the agreement is quite
good, taking into account the fact that we have averaged over 2000
realizations of the disordered versus 10000 in \cite{engel}and 40000 in \cite
{mp}. In doing a PIMC we did not have enough computer time to average over a
larger number of realizations.

In Fig. 4 we show the results of the Monte Carlo simulations for the mean
square displacement, for the case of long ranged correlated potential. For
comparison we show the results obtain from the analytical solution reported
in I. As in the classical case we observe that when quantum effects are
turned on, the RS symmetric solution gives an unphysical result below $%
T_c(\kappa )$, down to a certain critical $\kappa$. The RSB solution gives
rise to a flat behavior of $\left\langle \left\langle {\bf x}^2\right\rangle
\right\rangle _R\ $below $T_c(\kappa )$. The actual results of the QMC show
that the function is indeed monotonic, but no sharp transition is observed,
and it continues to decrease at all temperatures. A sharp transition is only
expected at $N=\infty $, whereas the simulation has been carried out at $N=1$,
where $N$ is the number of spatial dimensions. The variational
approximation also gives rise to a sharp transition at all 
dimensions, much like the large-$N$ result. As $\kappa \ $decreases,
tunneling increases and the glassines of the system decreases as the
particle is able to tunnel across potential barriers. This is evident in the
analytical solution by the decrease of $T_c(\kappa )$ with decreasing
$\kappa $, until there is no longer any transition.

     In this simulation we found that the statistical errors are 
dominated by the sample to sample fluctuations of   $\left\langle
 {\bf x}^2\right\rangle\ $ (The discussion of this quantity is in the 
next paragraph.).  For this reason, the error bars in Fig.4 are given by 
\begin{eqnarray}
 \pm 2 { 
  \left(
  {\left\langle 
        \left\langle {\bf x}^2\right\rangle 
        \left\langle {\bf x}^2\right\rangle 
\right \rangle}_R
 -{\left\langle \left\langle {\bf x}^2\right\rangle \right\rangle_R ^2}
 \right)}^{1/2}
 /\sqrt{(\mbox{number of samples})}  ,
\end{eqnarray}
with the number of samples being $2000$ in our case.  
  We give only the error bars for the case of $\kappa=100$ and 
$\kappa=0.1$ otherwise the figure would become too messy. The error
bars increase with increasing $\kappa$.  Also, in Table 2 we give the value of 
 $\left\langle \left\langle {\bf x}^2\right\rangle\right\rangle _R\ $
 for the first and second 1000 samples for $\kappa$=0.2, 0.3, and 1.0.
 We found that all the values of  $\left\langle \left\langle {\bf 
x}^2\right\rangle\right\rangle _R\ $ are within the error bars for those 
values of $\kappa$ which is very reasonable.

What is the signature of this effect in the simulations? In order to see
this effect we measured the sample to sample fluctuations of $\langle {\bf x}%
^2\rangle $. This is a direct measure for the glassiness of the system. In
Fig. 5 we depict the sample to sample fluctuation of $\langle {\bf x}%
^2\rangle $ and observe that for small enough $\kappa $ the function becomes
flat with decreasing temperature which signals the fact that quantum effects
wipe out the glassy behavior. This figure is to be compared with the Fig. 2
obtained from the variational approximation.

We also have to notice that the large sample to sample fluctuation of $%
\langle {\bf x}^2\rangle $ in the low temperature regime also indicates that
the error bars on our graphs for $\langle {\bf x}^2\rangle $ are relatively
large in the glassy phase. In fact the uncertainty of our calculation from
only 2000 samples in not enough to make a highly precise quantitative
description of the behavior of this system. Nevertheless, for a qualitative
study of this system at low temperature our simulation gives a clear picture.

Similar results were obtained for the case of short ranged correlated
potential. In Fig. 6 we display the PIMC results for the mean square
displacement together with the results obtained in I from the analytical
calculation for the RS solution. The analytical solution has been obtain for
power correlations with $\gamma =3/2$ , but from the variational
approximation any faster falling correlation should give similar results.

In Fig. 7 we show the sample to sample fluctuations of $\langle {\bf x}%
^2\rangle $ for the short ranged case, as obtained from the QMC simulations
to be compared with Fig. 3 obtained the analytical variational calculation.
Again we observe the reduction of the glassines of the system with the
increase of the quantum effects.

To conclude, we observe the relatively good agreement between the QMC
simulation and the results of the variational calculation. Some of the
deviations are of course due to the variational approximation, but they are
also due to the fact that in the simulation we averaged only over 2000
realizations. The results show that the sample to sample fluctuations of the
mean square deviation of the particle from the origin are a good measure for
the glassiness of the system which decreases with increasing quantum effects
(increase of $\hbar ^2/m)$. Thus the qualitative similarity with the phase
diagram of the quantum spin glass in a transverse field is established.

\acknowledgements
We thank the Pittsburgh Supercomputer Center for support under grant No.
DMR950018P. Y.Y.G. thanks the Weizmann Institute for support as a Michael
Visiting Professor, where part of this work has been done.

\newpage
\appendix
\section{}

For the temperature range where the RS solution is valid, it has been shown in
Appendix B of I how to obtain the numerical solution of the self consistent
equations (4.15) and (4.16) of I. When replica symmetry is broken, in the
case of long-ranged correlated potential, the equations for self energy
matrix are given in eqs. (5.3)-(5.8) of I. The solution of equation (5.6) of
I is obtained analytically there, and is presented in equations (5.9)-(5.11)
which are reproduced below.

The off diagonal elements parametrized by the Parisi variable $z$ are given
by: 
\begin{eqnarray}
\tilde s(z)=\left\{ 
\begin{array}{ll}
{\frac 32}Az_1^2 & \hspace{0.5 in}0<z<z_1\nonumber \\ 
{\frac 32}Az^2 & \hspace{0.5 in}z_1<z<z_2\nonumber \\ 
{\frac 32}Az_2^2 & \hspace{0.5 in}z_2<z<1
\end{array}
\right.  \label{rsb}
\end{eqnarray}
with 
\begin{eqnarray}
A &=&(2/3)^3g^2\beta ^3  \label{A} \\
z_1 &=&{\frac 32}g^{-2/3}\mu ^{1/3}\beta ^{-1}  \label{z1}
\end{eqnarray}
and $z_2$ is the solution of the equation 
\begin{eqnarray}
{\frac 12}\ \beta \ A\ a_R\ z_2^4\ +\ z_2\ -\ {\frac 34}=0\ .  \label{z2}
\end{eqnarray}
where

\begin{eqnarray}
a_R(\beta ,m,\mu ,g)=a_0+b_0(\beta ,\kappa ,\mu ,g),  \label{ar}
\end{eqnarray}
with 
\begin{eqnarray}
b_0(\beta ,\kappa ,\mu ,g)={\frac 2\beta }\sum_{\omega \neq 0}{\frac 1{%
m\omega ^2+\mu -\tilde s_d(\omega )}}.  \label{b0}
\end{eqnarray}
We integrate $\tilde s$ over $z$ to get $\int_0^1dz\ \tilde s(z)$ and
substitute in eq. (5.7) of I:

\begin{equation}
\tilde s_d(\omega )=\int_0^1dz\ \tilde s(z)-{\frac 2\hbar }\int_0^Ud\zeta \
(1-e^{i\omega \zeta })\ f^{\prime }\left( {\frac 2\beta }\sum_{\omega
^{\prime }\neq 0}\ {\frac{1-e^{-i\omega ^{\prime }\zeta }}{m\omega ^{\prime
}{}^2+\mu -\tilde s_d(\omega ^{\prime })}}\right)
\end{equation}

Then we have a set of nonlinear equations for $\tilde s_d(\omega )$'s which
are solved for up to 20 non-zero Matsubara frequencies ($l=-10\cdots 10$, in
eq. \ref{omega}). The solution is then used (utilizing eq.(5.4) of I) to
calculate the sample to sample fluctuation of $\langle {\bf x}^2\rangle $ as
given in eq. \ref{x22b}, and is depicted in Fig. 2.

For the case of short-ranged correlation, we seek a solution with one step
replica symmetry breaking, i.e., 
\begin{eqnarray}
\widetilde{s}(z)=\left\{ 
\begin{array}{ll}
s_0 & \hspace{0.5 in}0<z<z_c\nonumber \\ 
s_1 & \hspace{0.5 in}z_c<z<1
\end{array}
\right.
\end{eqnarray}
and hence

\begin{eqnarray}
\lbrack \tilde s](z)=z\ \tilde s(z)-\int_0^zdz\ \tilde s(z).  \label{sbraket}
\end{eqnarray}
is given by 
\begin{eqnarray}
\lbrack \widetilde{s}](z)=\left\{ 
\begin{array}{ll}
0 & \hspace{0.5 in}0<z<z_c\nonumber \\ 
\Sigma & \hspace{0.5 in}z_c<z<1\nonumber
\end{array}
\right. ,
\end{eqnarray}
where $\Sigma =z_c(s_1-s_0)$. The breaking point and the order parameters $%
s_0$, $s_1$ are found from maximizing the variational free energy. The
equations are similar to the classical case \cite{engel}, the difference
between the classical and the quantum case enters again through the
renormalization of the constant $a_0\rightarrow a_R=a_0+b_0$ which enters
the correlation function of the random potential.

We thus obtain a set of self-consistent equations for the $\tilde s_d(\omega
)$'s and also $\tilde s(z)$ (or $s_0$ and $s_1$) at the same time and the
Newton-Raphson method can be applied. When replica symmetry is not broken we
simply get $u_c=1$. The solution is then used (utilizing eq.(5.4) of I) to
calculate the sample to sample fluctuation of $\langle {\bf x}^2\rangle $ as
given in eq. \ref{x22b}, and is depicted in Fig. 3.

\newpage
Table Caption: 
\begin{itemize}

\item

Table 1: Example of different parameters in the QMC for the
long ranged correlated potential.

\item
Table 2: Example of $\left\langle \left\langle{\bf x}^2 \right\rangle
\right\rangle _R$  from two sets of 1000 realizations of the long
ranged correlated random potential in the QMC. 

\end{itemize}

\begin{tabular}{|c|c|c|c|c|}
\hline
$\kappa =100$ &  &  &  &  \\ \hline
t & dx & $a_0$ & $a_0^{\prime }$ & $a_1$ \\ \hline
3.0 & 0.02 & 2.30 & 2.0 & 0.03 \\ 
2.0 & 0.02 & 2.00 & 1.3 & 0.04 \\ 
1.0 & 0.03 & 1.70 & 0.4 & 0.04 \\ 
0.9 & 0.03 & 1.50 & 0.4 & 0.04 \\ 
0.8 & 0.03 & 1.20 & 0.3 & 0.05 \\ 
0.7 & 0.03 & 1.00 & 0.2 & 0.05 \\ 
0.6 & 0.04 & 0.90 & 0.15 & 0.06 \\ 
0.5 & 0.04 & 0.90 & 0.12 & 0.06 \\ 
0.4 & 0.05 & 0.80 & 0.10 & 0.07 \\ 
0.3 & 0.05 & 0.80 & 0.09 & 0.07 \\ 
0.2 & 0.06 & 0.70 & 0.08 & 0.08 \\ \hline
$\kappa =0.2$ &  &  &  &  \\ \hline
t & dx & $a_0$ & $a_0^{\prime }$ & $a_1$ \\ \hline
3.0 & 0.30 & 1.45 & 1.45 & 0.80 \\ 
2.0 & 0.30 & 1.05 & 1.05 & 0.90 \\ 
1.0 & 0.35 & 1.00 & 1.00 & 0.95 \\ 
0.9 & 0.37 & 0.95 & 0.80 & 0.85 \\ 
0.8 & 0.48 & 0.90 & 0.70 & 0.85 \\ 
0.7 & 0.59 & 0.90 & 0.60 & 0.85 \\ 
0.6 & 0.60 & 0.85 & 0.50 & 0.80 \\ 
0.5 & 0.71 & 0.85 & 0.45 & 0.80 \\ 
0.4 & 0.82 & 0.80 & 0.40 & 0.80 \\ 
0.3 & 0.82 & 0.80 & 0.35 & 0.66 \\ 
0.2 & 0.82 & 0.70 & 0.20 & 0.45 \\ \hline
\end{tabular}
\vskip 1cm

\begin{tabular}{|c|c|c|}
\hline
$\kappa=1.0$ &  & \\  \hline
t&  $\left\langle \left\langle{\bf x}^2 \right\rangle \right\rangle
_R$, first 1000 samples  & 
     $\left\langle \left\langle{\bf x}^2 \right\rangle \right\rangle
     _R$, second 1000 samples    \\ \hline 
3.0 & 4.185  &   4.058  \\
2.0 & 3.396  &   3.281   \\
1.0 & 2.767  &   2.662  \\
0.8 & 2.672  &   2.574  \\
0.6 & 2.601  &   2.497  \\
0.4 & 2.552  &   2.437  \\ 
0.2 & 2.519  &   2.404  \\   \hline
$\kappa=0.3$ &  & \\  \hline
t&  $\left\langle \left\langle{\bf x}^2 \right\rangle \right\rangle
_R$, first 1000 samples  & 
     $\left\langle \left\langle{\bf x}^2 \right\rangle \right\rangle
     _R$, second 1000 samples    \\ \hline 
 3.0 & 4.237 & 4.112  \\
 2.0 & 3.465 & 3.352  \\
 1.0 & 2.857 & 2.755  \\
 0.8 & 2.774 & 2.670  \\
 0.6 & 2.701 & 2.603  \\
 0.4 & 2.654 & 2.552  \\
 0.2 & 2.610 & 2.498   \\    \hline
$\kappa=0.2$ &  & \\  \hline
t&  $\left\langle \left\langle{\bf x}^2 \right\rangle \right\rangle
_R$, first 1000 samples  & 
     $\left\langle \left\langle{\bf x}^2 \right\rangle \right\rangle
     _R$, second 1000 samples    \\ \hline 
 3.0 & 4.276 & 4.148  \\
 2.0 & 3.514 & 3.402  \\
 1.0 & 2.925 & 2.824  \\
 0.8 & 2.846 & 2.745  \\
 0.6 & 2.784 & 2.683  \\  
 0.4 & 2.734 & 2.636  \\
 0.2 & 2.677 & 2.577  \\    \hline
\end{tabular}

\newpage
{\large {\bf Figure Captions}} \\

\begin{itemize}
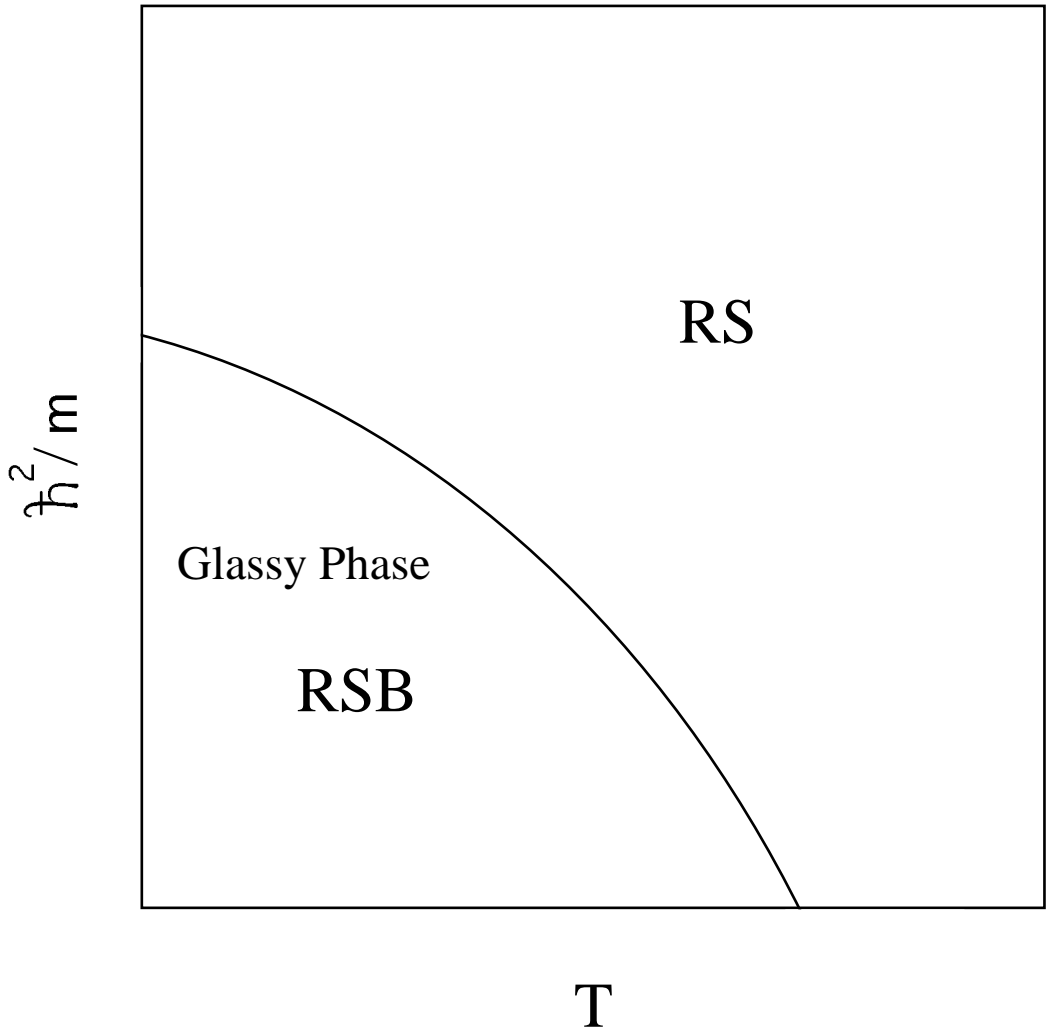
\item  Fig 1. Schematic phase diagram of a quantum particle in a random
potential plus an harmonic potential, as obtained from the variational
calculation derived in I.

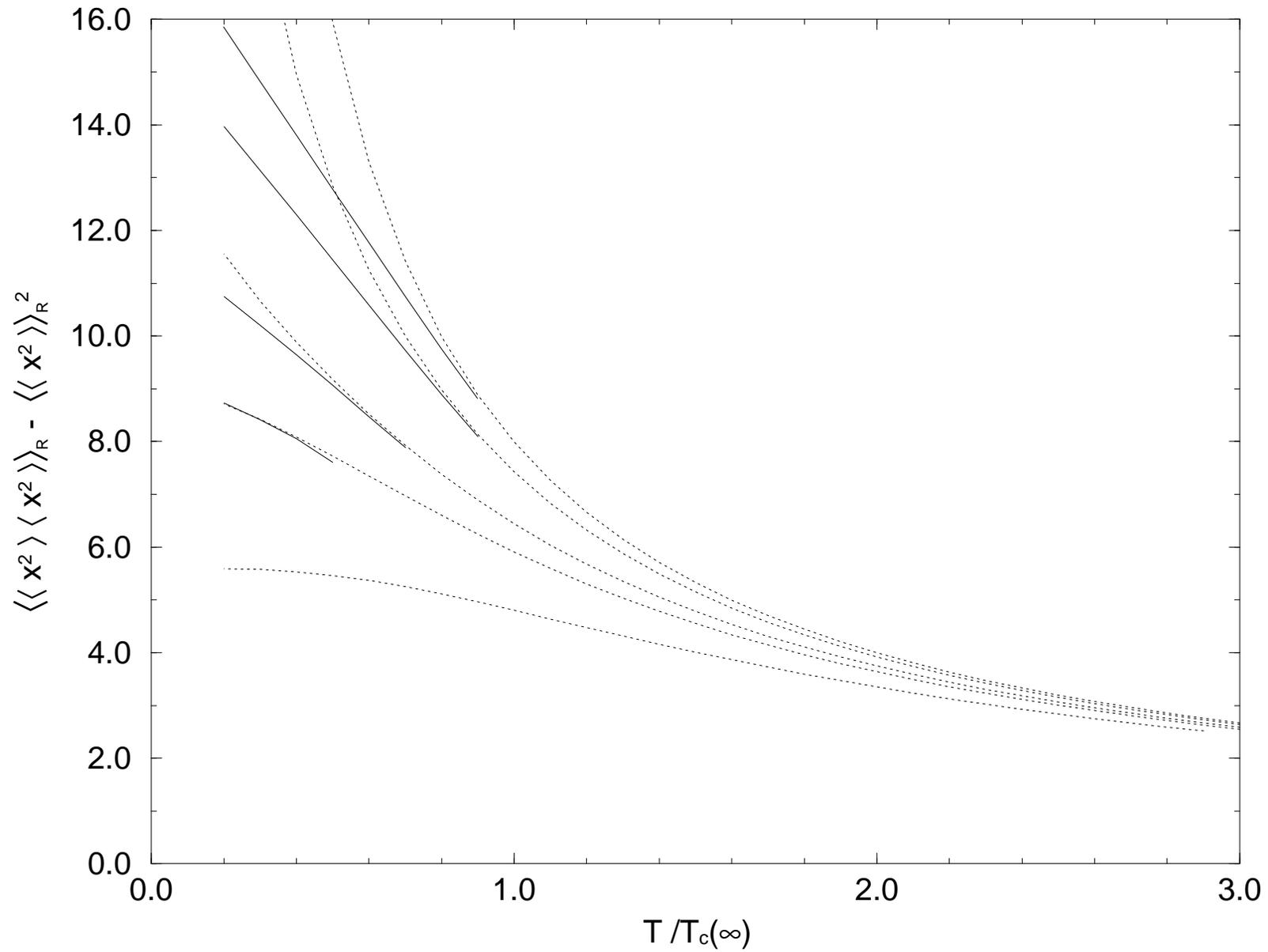
\item  Fig 2. Plot of sample to sample fluctuation of $\left\langle
\left\langle {\bf x}^2\right\rangle \right\rangle _R$ for long-ranged
correlated disorder from numerical solution. Dashed lines are solutions
assuming replica symmetry. Solid lines are RSB solutions. From top to
bottom: $\kappa =100$, 1.0, 0.3, 0.2, 0.1.

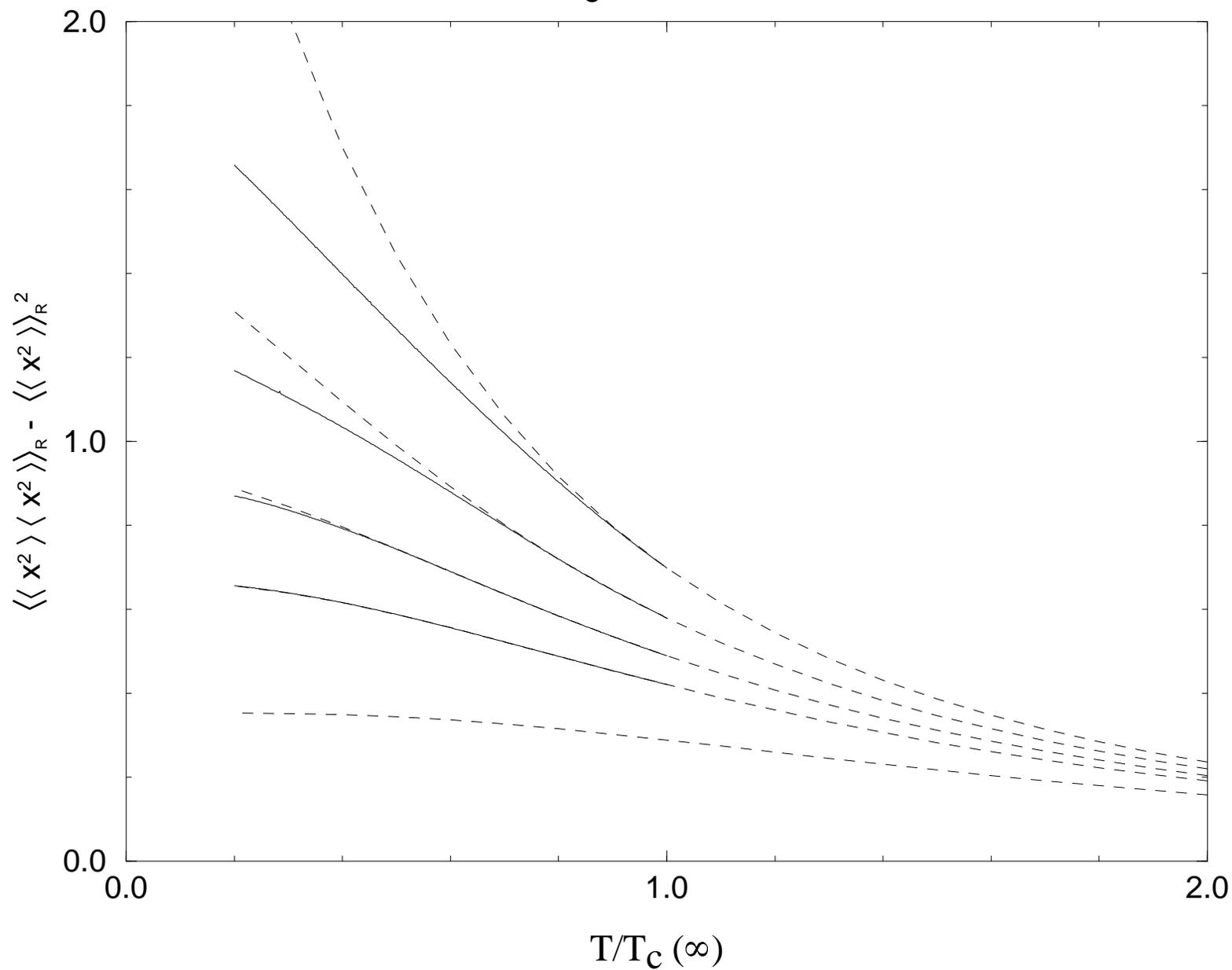
\item  Fig 3. Plot of sample to sample fluctuation of $\langle {\bf x}
^2\rangle $ for short-ranged correlated disorder from the numerical
solution. Dashed lines are solutions assuming replica symmetry. Solid lines
are 1-step RSB solutions. From top to bottom: $\kappa =$100, 6, 3, 2, and 1.

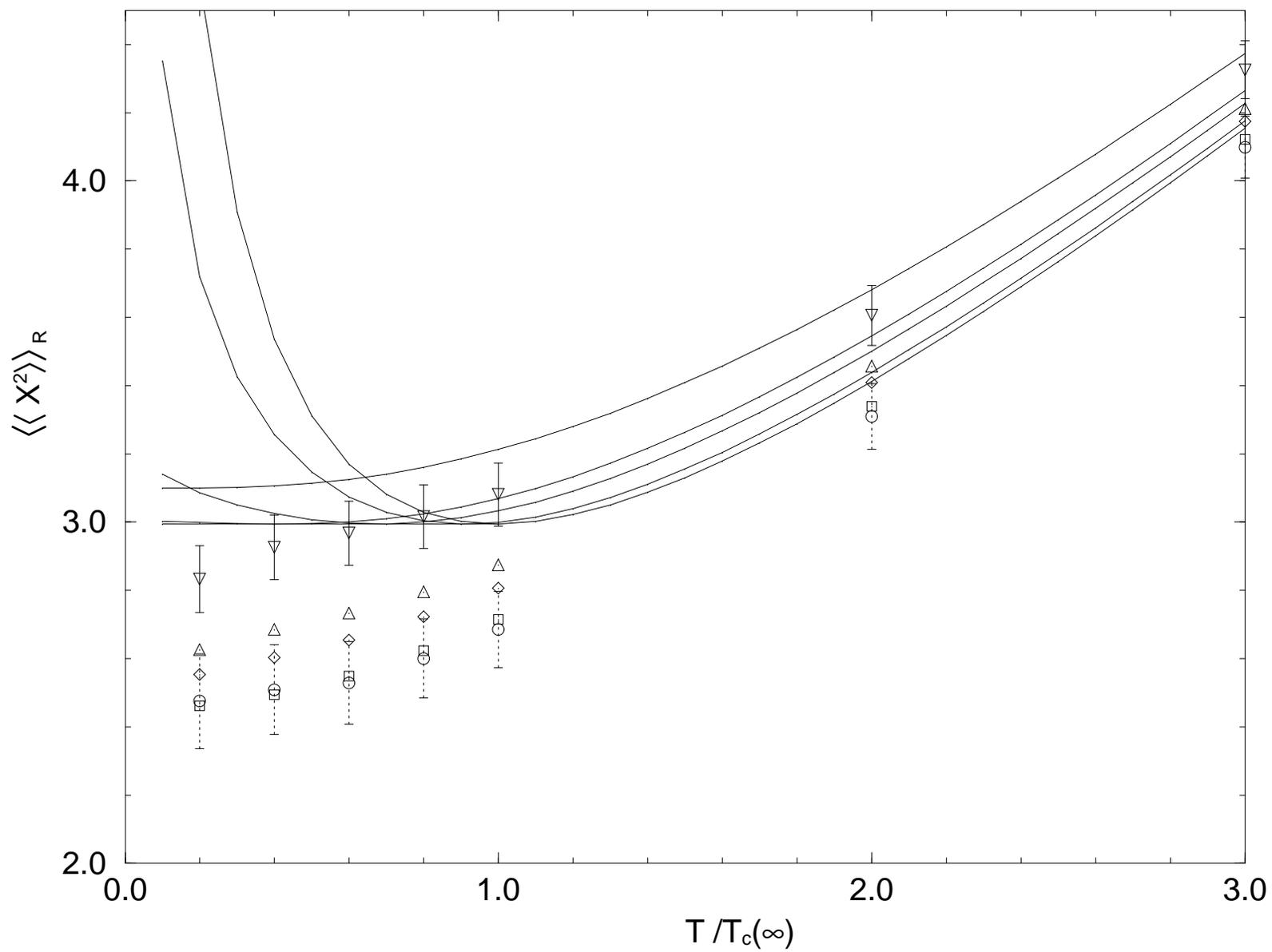
\item  Fig 4. Plot of $\left\langle \left\langle {\bf x}^2\right\rangle
\right\rangle _R$ vs. $T/T_c(\infty )$ for long-ranged correlated disorder.
Solid lines are numerical solutions, obtained in I, assuming replica
symmetry. From top to bottom: $\kappa =$0.1, 0.2, 0.3, 1.0, 100. Data points
are Monte Carlo simulation for $\kappa =100$ ( circles), $\kappa =1.0$
(squares), $\kappa =0.3$ (diamonds), $\kappa =0.2$ (up triangles), and $%
\kappa =0.1$ (down triangles). Each point is averaged over 2000 samples.
Error bars indicate the statistical errors for the cases of $\kappa
=100$ (dotted line) and $\kappa = 0.1$ (solid line).

\item  Fig 5. Plot of sample to sample fluctuation of $\langle {\bf x}
^2\rangle $ for long-ranged correlated disorder from Monte-Carlo simulation.
Data points are for $\kappa =100$ ( circles), $\kappa =1.0$ (squares), $%
\kappa =0.3$ (diamonds), $\kappa =0.2$ (up triangles), and $\kappa =0.1$
(down triangles). Each point is averaged over 2000 samples.

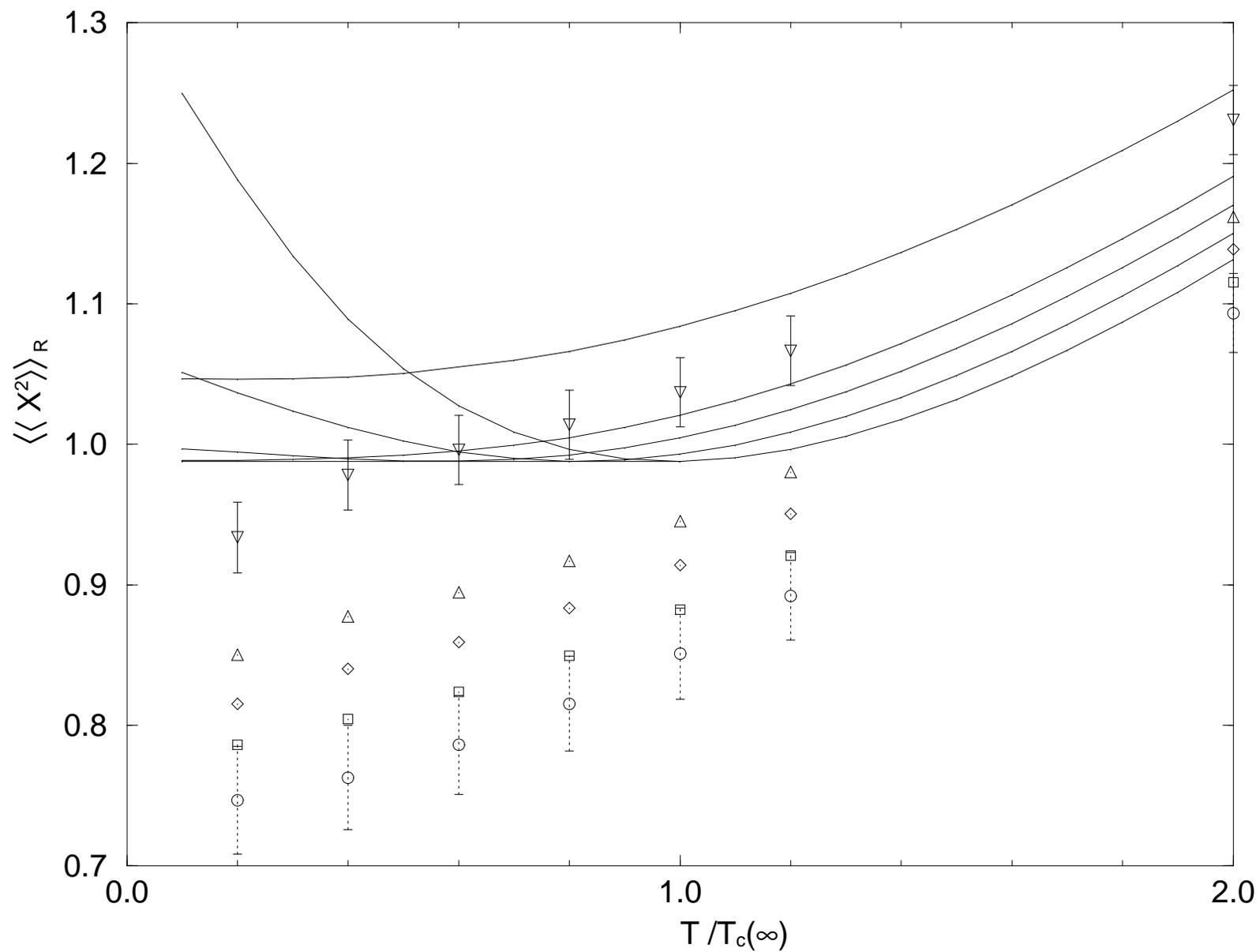
\item  Fig 6. Plot of $\left\langle \left\langle {\bf x}^2\right\rangle
\right\rangle _R$ vs. $T/T_c(\infty )$ for short-ranged correlated disorder.
Solid lines are numerical solutions, obtained in I, assuming replica
symmetry. From top to bottom: $\kappa $=1, 2, 3, 6, 100. Data points
are Monte Carlo simulation for $\kappa 
=100$ (circles), $\kappa =6$ (squares), $\kappa =3$ (diamonds), $\kappa =2$
(up triangles), $\kappa =1$ (down triangles). Each point is averaged over
2000 samples.  Error bars indicate the statistical errors for the
cases of $\kappa =100$ (dotted line) and $\kappa = 1$ (solid line).

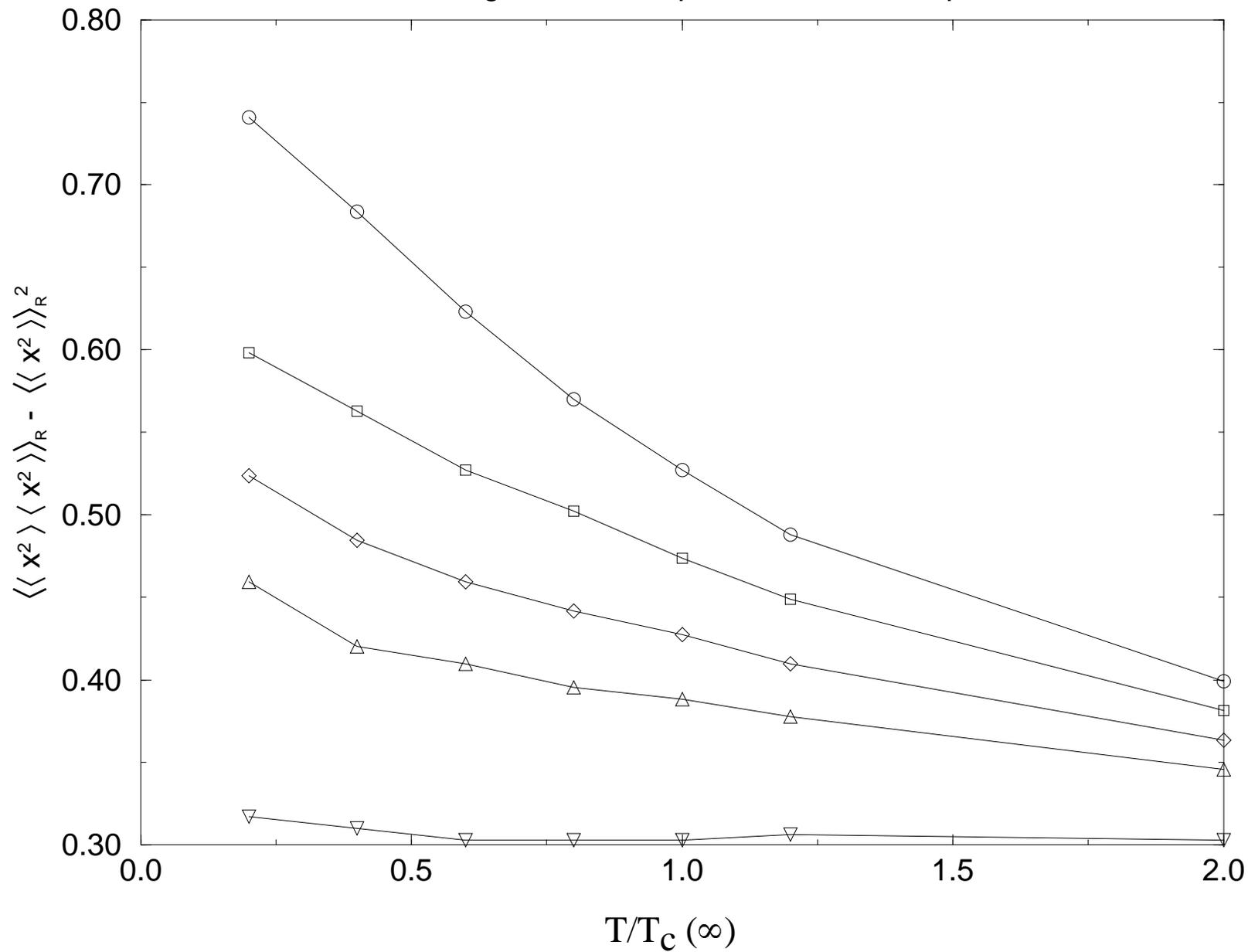
\item  Fig 7. Plot of sample to sample fluctuation of $\langle {\bf x}
^2\rangle $ for short-ranged correlated disorder from Monte Carlo simulation.
Data points are for $\kappa =100$ (circles), $\kappa =6$ (squares), $\kappa
=3$ (diamonds), $\kappa =2$ (up triangles), $\kappa =1$ (down triangles).
Each point is averaged over 2000 samples.
\end{itemize}

\end{document}